\documentclass[letterpaper]{article}
\usepackage{aaai}
\usepackage{times}
\usepackage{helvet}
\usepackage{courier}
\usepackage{graphicx}
\usepackage{amsmath,amssymb}
\usepackage{subfig}
\usepackage{algorithm}
\usepackage{algorithmic}
\usepackage{multirow}
\usepackage{array}
\usepackage{balance}
\usepackage{url}
\begin{document}

\title{Competition and Success in the Meme Pool: a Case Study on Quickmeme.com}
\author{Michele Coscia\\
CID - Harvard Kennedy School\\
79 JFK St, Cambridge, MA, US, 02139\\
michele\_coscia@hks.harvard.edu
}

\maketitle
\begin{abstract}
The advent of social media has provided data and insights about how people relate to information and culture. While information is composed by bits and its fundamental building bricks are relatively well understood, the same cannot be said for culture. The fundamental cultural unit has been defined as a ``meme''. Memes are defined in literature as specific fundamental cultural traits, that are floating in their environment together. Just like genes carried by bodies, memes are carried by cultural manifestations like songs, buildings or pictures. Memes are studied in their competition for being successfully passed from one generation of minds to another, in different ways. In this paper we choose an empirical approach to the study of memes. We downloaded data about memes from a well-known website hosting hundreds of different memes and thousands of their implementations. From this data, we empirically describe the behavior of these memes. We statistically describe meme occurrences in our dataset and we delineate their fundamental traits, along with those traits that make them more or less apt to be successful.
\end{abstract}

\section{Introduction}
Social media are virtual communities present on the Web that allow people to create, share, exchange and comment on pieces of content among themselves. Social media can be of different types: social networks, whose main aim is to allow people to keep in touch with hundreds of friends, like Facebook or Google+; social bookmarking websites, that allow users to share links to interesting content present on the Web, like Reddit or Digg; blogging platforms, where the user itself is posed at the center of the content creating process, like Wordpress or Blogger; and many more. The defining characteristic of social media is the many-to-many communication: the users are at the same time producers and consumers of information and knowledge.

In social media, a popular concept is the one of ``Internet Meme''. A ``Meme'' is defined as the simplest cultural unit that can spread from one mind to another \cite{selfishgene}. A particular tune or a given rhetoric figure are examples of memes. An internet meme is a meme that spreads through the internet. Internet memes carry an additional property that ordinary memes do not. While preserving all the characteristics of ordinary memes, due to being spread through the internet, internet memes leave a footprint that is traceable and analyzable. Meme spreading through other media are not so easily traceable. For this reason, several researchers already studied internet memes in social media like Twitter \cite{faloutsos-meme} or the blogosphere \cite{tracking-meme1}.

However, most studies of internet memes share a common starting point: they observe the usage of memes by users who are interacting in a network. The main focus is on the interactions between the users and the influence of the topology of the network itself in the meme spreading process. In other words, these studies are not inquiring about the characteristics and the dynamics of memes per se, but the characteristics of the environment in which memes live, i.e. the social network that lies underneath social media. A meme is studied only in terms of its reaction to this environment. For this reason, we say that these works are studying the ecology of internet memes. We chose an alternative approach by focusing on the description of the characteristics and the behavior of some internet memes, independently from the networks they live in, like in \cite{memestudy}.


In this paper we focus mainly on the analysis of internet meme data from  Quickmeme\footnote{\url{http://www.quickmeme.com/}}. Quickmeme is a website mainly used by social bookmarking users to create memes and share them on a social bookmarking website (Quickmeme was created by Reddit users to have a platform where to create and share memes on Reddit itself). Quickmeme is an instrument to track memes because it registers without any ambiguity the meme used, its rating and the moment in time in which the meme has been created. We are aware that Quickmeme is not containing most internet memes, nor all kinds of internet memes. However, by focusing on this easily traceable source we are able to study a portion of internet memes in a controlled environment, allowing us for higher quality results that may be generalized to internet memes in general, in a second time.

The aim of the paper is twofold, focused on meme competition and meme success. First, we want to prove that the memes we are tracking via the Quickmeme website have the fundamental characteristics to be called ``memes''. To do so, we need to prove that they are interacting, as variation and heredity are noted to be two fundamental characteristics of memes \cite{selfishgene}. Their interactions can take the form of competition and collaboration, where memes compete for the attention of the users and, in doing so, they can also cooperate resulting in higher success ratios.

Second, we want to study the characteristics that make them successful memes in Quickmeme. Given some quantitative characteristics of the memes, we want to understand how these characteristics are influencing their chances of being successful memes. The difference with the rest of the internet meme literature is that we are not considering any social or network effect when studying meme success. The typical question of those studies breaks down to: ``what are the characteristics of a social network which maximize information spreading?''. Instead, we think that the success of a meme may be influenced by where it appears for the first time, but ultimately it also has to have some characteristics that make it more apt to survive.

To sum up, this paper makes the following contributions:

\begin{itemize}
\item We are providing a study of the intrinsic characteristics of some memes, without focusing the network effect behind them, thus providing useful insights to better understand cultural dynamics, instead of social dynamics;
\item We provide insights about a novel source of data for memetics studies, by being able to detect and study memes in social media, using a data source like Quickmeme that provides high quality data about meme popularity;
\item We are able to detect competition and collaboration in the meme pool of Quickmeme, and to study the characteristics of successful memes present in the website.
\end{itemize}

The remainder of the paper is organized as follows. In Section \ref{sec:related} we review the related literature on internet memes, information spread and memetics. We present our data source and the data cleaning process in Section \ref{sec:data}. Sections \ref{sec:competition} contains the Quickmeme analysis methodology and the evidences for meme relationships (to detect competition and collaboration) in our data source. In Section \ref{sec:success} we try to define the characteristics that lead to meme success in our dataset. Section \ref{sec:conclusions} concludes the paper by pointing out future developments of this work.

\section{Related Work}\label{sec:related}
This paper is connected to several different tracks of research: the study of how the social network and the social media allows memes to spread (what we call ``meme ecology''); the study of social media in general and memetics.

In the Introduction we stated that there are several works studying internet memes. For example, \cite{faloutsos-meme} studies the dynamics of competitions of memes spreading in two distinct social networks on different social media (Facebook and Twitter). Also \cite{competition-meme1} addresses the problem of competition among memes, assuming that each user can follow only a handful of memes at the same time. Meme and information spread is also a problem definition addressed in computer science with different methodological approaches: works like \cite{gurumine}, \cite{vir} and \cite{ida09} provide algorithms and frameworks to analyze how information spreads in a network of people. While those papers examine different cascade information spreads as independent, in \cite{lesko-contagion} different spread events on the same network are analyzed at the same time, as different memes influence each other while trying to span on the same set of minds. Other computer science tools help us tracking memes over the Web \cite{tracking-meme1}. As an application, \cite{spreading-meme1} deals with cooperative behavior. 

We do believe that the social network analysis behind meme spreading is interesting, however it leaves undescribed the fundamental characteristics of the memes themselves. While studying a system, it is important to know how the parts interact, but also how they function themselves, to have a better picture of what actually is happening in the world. Our paper tries to provide some contribution exactly in this last aspect: we do not consider network analysis, only the memes themselves. Closer to our work is \cite{memestudy}, but here the author does not address the issues of collaboration and competition in internet memes.

Internet memes spread over social media websites. In computer science, many researchers have addressed the problem of describing the dynamics of user behavior in social media websites. The topics touched include the emergence of conventions in online social networks \cite{cha}, how to select the critical features in the amount of data generated in these websites \cite{liu}, the privacy aspects \cite{privacy}, the ``follow'' and friendship dynamics in Twitter \cite{tweet} and \cite{lada}, and many more.

Finally, there are in literature also some publications about memetics, the proposed branch of science that should study memes. Some of the pioneering works are \cite{hofstadter}, \cite{brodie2004} and \cite{lynch}. Our work is different from these examples in literature as we are focused particularly on internet memes. Our approach is more data driven, as internet memes are more easily traceable as they leave a measurable footprint, this makes our paper a further contribution w.r.t these works.

\section{The Data}\label{sec:data}
\begin{figure}
\centering
\subfloat[Socially Awkward Penguin]{\includegraphics[width=.475\columnwidth]{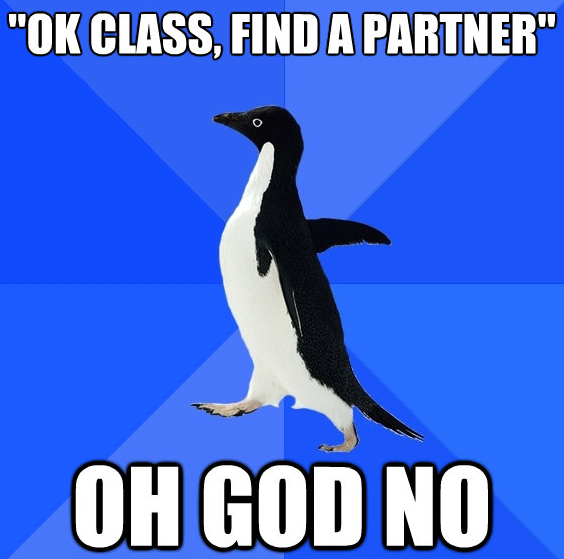}}\quad
\subfloat[Philosoraptor]{\includegraphics[width=.475\columnwidth]{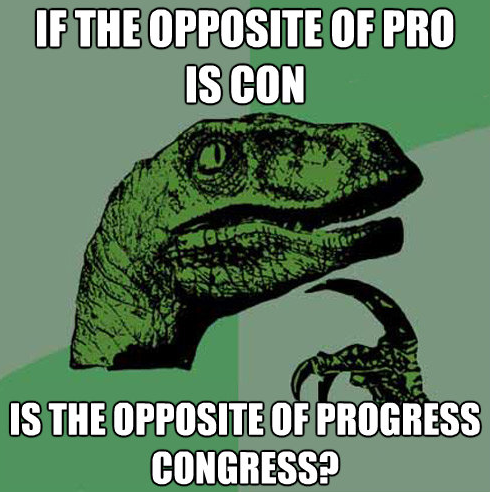}}
\caption{Two implementations of memes from Quickmeme.}
\label{fig:meme-example}
\end{figure}

As stated in the Introduction, social media provide many sources for meme analysis and we chose to focus on the website Quickmeme. There are other kinds of internet memes and different dynamics for their creation, thus we will only describe the meme types present in Quickmeme.

Quickmeme works in the following way: the website provides a system to create memes and then to create particular implementations of any other meme. We provide a couple of examples in Figures \ref{fig:meme-example}(a-b). Figure \ref{fig:meme-example}(a) shows an implementation of the meme ``Socially Awkward Penguin''. The meme is a picture with a left facing penguin on blue field. The socially awkward penguin is used to make jokes about anything related to social clumsiness: users use the template to describe a social situation where they misbehaved or they did not know how to properly react. Figure \ref{fig:meme-example}(b) is the ``Philosoraptor'' meme and it is depicted as a Velociraptor in a philosophical pose in green field. It is commonly used for rhetoric questions and puns sounding philosophically deep.

From these examples, we can define a meme as a combination of a picture and a tacit concept linked to the picture. We refer to a meme with the symbol $m$. An implementation of a meme is the picture and a particular humorous caption added to it following the tacit concept of the meme. An implementation of a meme $m$ is referred as $i^n_m$, so $m = \{i^1_m, ... , i^n_m\}$ is the collection of all $n$ implementations of meme $m$.

Quickmeme provides a suitable data source for our study for different reasons. First, users are versatile in their meme use: they combine memes, they make them evolve with different pictures or concepts, they use one meme against others. For example, from the socially awkward penguin, users created the ``Socially Awesome Penguin'', the ``Socially Average Penguin'' and tens more. These dynamics of evolution, competition and collaboration are not much different from the same dynamics observed in the gene pool.

Second, the website provides a scoring system, that can be used as a proxy of a meme's ``success'' in the meme pool. Without the need of an account, anybody can cast a vote saying that a particular meme was particularly funny, recognizable or otherwise remarkable. When voting, the users have three choices: ``awesome'' meme (that adds $2$ to the total rating of the meme), ``average'' meme (that adds $1$) and ``bad'' meme (that removes $1$ from the total rating).

In Quickmeme, a meme implementation can be ``Featured'' if it is sufficiently popular. We crawled the 499 memes with at least one featured implementation present in the website on October 15th, 2012, restricting our crawl to only the implementations of the memes created since October 9th, 2011. We chose to restrict to memes with featured implementations as they are more visible, thus they obtain more votes and generate more data points for our study. We  downloaded $178,801$ meme implementations in total.

\begin{figure}
\centering
\includegraphics[width=.95\columnwidth]{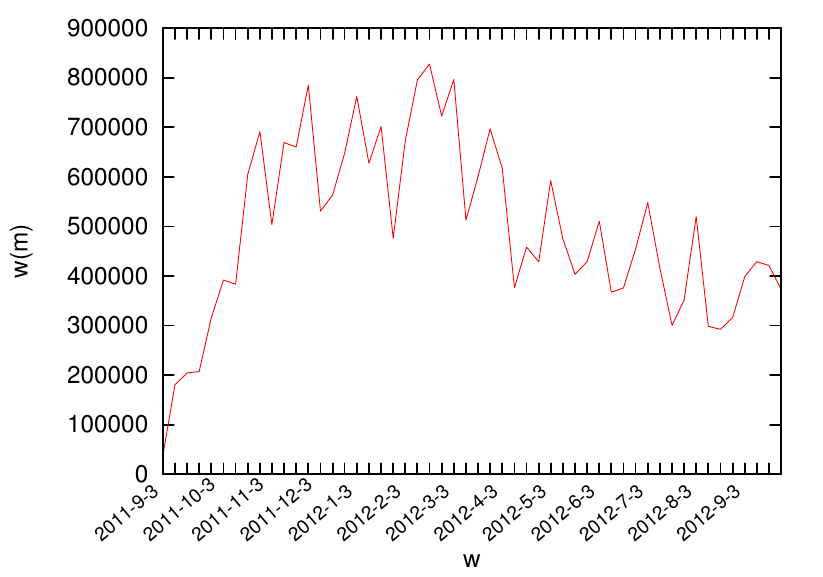}
\caption{Temporal distribution of meme ratings.}
\label{fig:stats1}
\end{figure}

Figure \ref{fig:stats1} shows the sum of ratings for all the memes. We can see that Quickmeme during the first weeks saw a growth in positive ratings cast. It reached its peak, in number of ratings, during the first quarter of 2012 starting a slow decline.

\begin{figure}
\centering
\includegraphics[width=.95\columnwidth]{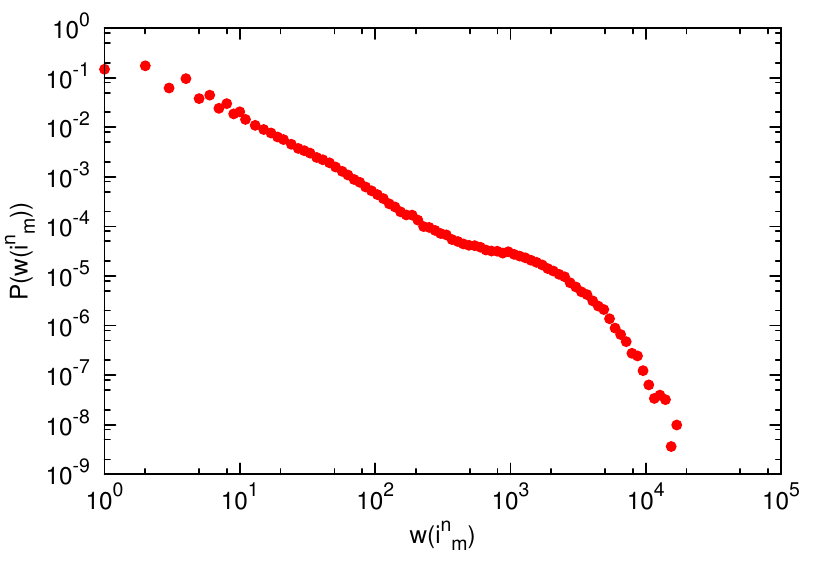}
\caption{Meme implementation rating distribution.}
\label{fig:stats2}
\end{figure}

In Figure \ref{fig:stats2} we show how many meme implementations $i^n_m$ obtained a particular rating score. We can identify a distribution roughly compliant with a power law: there are some very popular memes (spanning across five orders of magnitude) while the vast majority has a low rating. There are two noticeable deviations: a hunch around $500$-$2,000$ ratings and the exponential cutoff. While the second is present in many distributions, the first is more interesting. This is a ``Front page'' effect: popular memes are usually exposed in the front page of the website, so that they obtain extra ratings. This explains why more memes than expected tend to have a rating $>1,000$.

\section{Detecting Meme Relationships}\label{sec:competition}
The aim of this section is to show that, in Quickmeme, the evolution of popularity of a meme is not dependent only on the meme itself, but it is influenced by the other memes. The influence may take place in two different ways: by means of \textit{competition} and by means of \textit{collaboration}. Competition is defined as a negative influence of one meme over another: the success in terms of ratings of one meme provokes lower ratings in another meme. Collaboration is the opposite effect. The influence may happen for many reasons (the memes are similar so a user with an idea for a caption must choose one of the two, one meme is used to criticize another meme for being useless or a new meme replaces an old meme), but our aim is simply to demonstrate that such influences exist.

To see whether the popularity of two memes is not independent, we look at the weekly total ratings of a meme. We refer to the total ratings of an implementation $i^n_m$ of meme $m$ in the first week as  $w_1(i^n_m)$. For each week, we sum all the ratings collected by the meme implementations of a meme, i.e $w_1(m) = \sum_{j=1}^{n} w_1(i^j_m)$ and $w(m) = \{w_1(m), w_2(m), ..., w_{53}(m)\}$ given that we collected data for 53 weeks. $w(m)$ is referred as ``meme rating vector''. If an increase in success of one meme tends to cause a fall in success of another, and viceversa, then we say that there is a competition between the two memes. A meme's success is determined by looking at its rating vector.

We discuss how we confront meme rating vectors via a null model definition in Section \ref{sec:nullmodel}. We then discuss some examples of collaboration in Section \ref{sec:collaboration} and we briefly present other instances of meme competition in Section \ref{sec:examples}.

\subsection{Null Model}\label{sec:nullmodel}
\begin{figure*}
\centering
\subfloat[Timeline of ratings]{\includegraphics[width=.95\columnwidth]{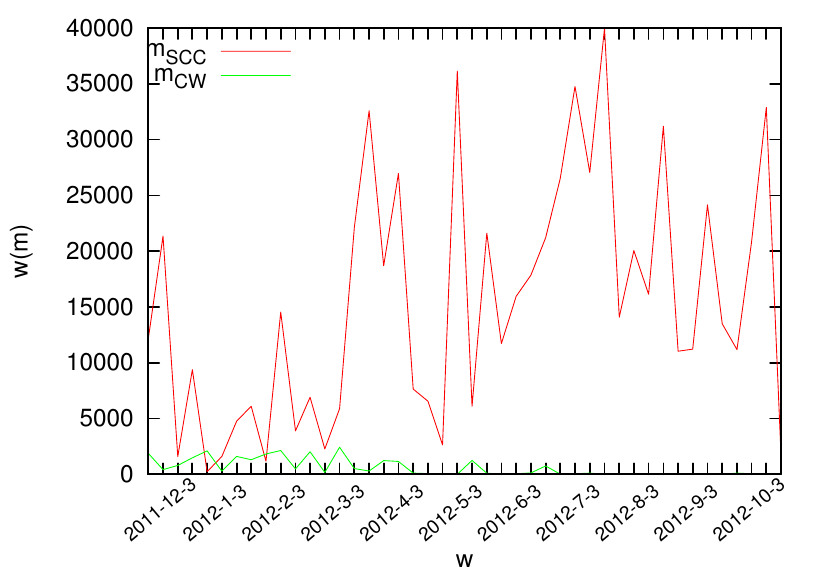}}\quad
\subfloat[Ratings vs null model]{\includegraphics[width=.95\columnwidth]{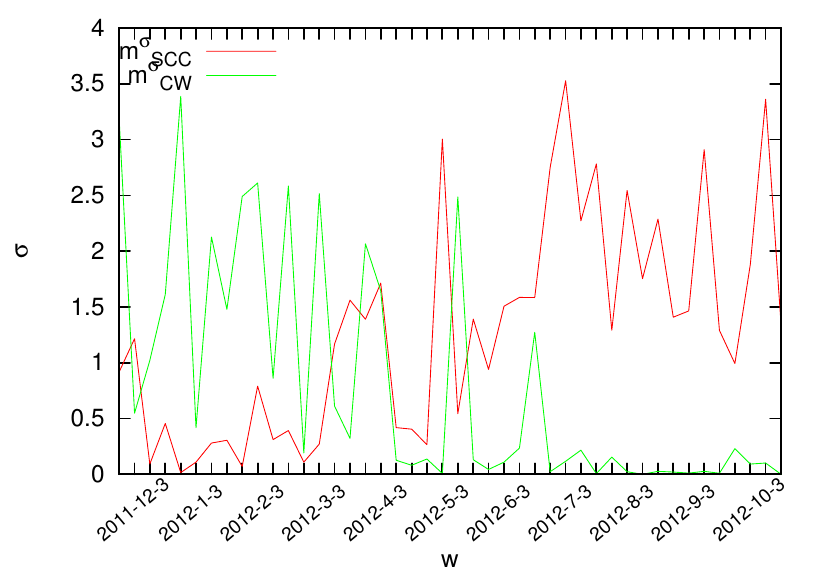}}
\caption{Temporal evolution of ratings and expected ratings of two competing memes.}
\label{fig:competition}
\end{figure*}

Two meme rating vectors may not be comparable for two reasons. To understand those reasons, we depict two meme rating vectors in Figure \ref{fig:competition}(a): in red we have the ratings evolution of the meme ``Sudden Clarity Clarence'' (SCC) while in green we have the ratings evolution of ``Courage Wolf'' (CW). As we can see, SCC is, \textit{ceteris paribus}, on average more popular than CW. As a result, we would find a competition even if there is none, as CW has always low ratings when SCC has high ratings regardless of their interactions.

The second reason is that two competing memes may rise together only because they became more popular for independent reasons or the Quickmeme website has a spike in its user traffic. In other words, fluctuations in ratings may be explained by meme and weekly website popularity, without necessarily meaning that the two memes are in competition or in collaboration. A null model will help us eliminate these meaningless fluctuations and generated a normalized meme popularity vector, as depicted in Figure \ref{fig:competition}(b).

The overall plan of action is the following:

\begin{itemize}
\item We generate a null model;
\item We generate normalized meme rating vectors by confronting the observed meme rating vectors with the expectations obtained from the null model;
\item We confront the normalized meme rating vectors to spot influences in meme popularity.
\end{itemize}

\subsubsection{Null Model Generation}
In the null model we want to assume there are no collaboration/competition relations, but we want to control for each meme's general popularity and the website popularity in a given week. This will be the baseline against which the observed data can be compared. To achieve this, we borrow a method from the ecology literature \cite{econull-1} \cite{econull-2}. We do this, because our problem has some similarities with the ecology models of species and ecosystems: memes are like animal species and their total ratings in one week is a quantitative measure of success of the species. The ecosystem of the species is represented by the users connected to the website through that particular week.

First, we store all meme rating vectors $w(m)$ in a meme matrix $M_{m,w}$. The meme rating vectors are the rows of the matrix, while the weeks are the columns. For example, the entry $M_{m,w}(i,j)$ contains the ratings of meme $m_i$ for the week $w_j$, or $w_j(m_i)$. Next, we create a randomized null version of $M_{m,w}$ in which each entry $w_j(m_i)$ is random, but we preserve the total sums of the rows (memes) and the columns (weeks) of the matrix. By testing the ratio between the observed values and the expected values according to this null model we have a quantification about how much a meme is unusually (im)popular, given its general popularity and given the popularity of Quickmeme that week.

We generated 100 of such null matrices using the vegan package for the R statistical software, that implements the Patefield algorithm \cite{matrix-null}. From our 100 null matrices we extracted three different versions of a null matrix. The first is $N_{m,w}$, the master null model, generated by averaging the cells of all 100 null matrices. Each cell $n_j(m_i) \in N_{m,w}$ is calculated using the following equation: $n_j(m_i) = \sum_{k = 1}^{100} \dfrac{w_j^k(m_i)}{100},$ where $w_j^k(m_i)$ is the value of cell $(i, j)$ in the $k$-th null matrix. The second and the third are respectively $N_{m,w}^{1}$ and $N_{m,w}^{2}$, constructed exactly like $N_{m,w}$, except that for $N_{m,w}^{1}$ we use the first 50 random matrices and for $N_{m,w}^{2}$ we use the remaining 50.

\subsubsection{Normalized Meme Rating Vectors}
We are now able to perform our data cleaning, to generate the normalized meme rating vectors that we have depicted in Figure \ref{fig:competition}(b). Each observed cell value $w_j(m_i) \in M_{m,w}$ is tested against its corresponding null model cell $n_j(m_i) \in N_{m,w}$. We then calculate each normalized value $\sigma_j(m_i)$ as follows:

\[
\sigma_j(m_i) =
\begin{cases}
 1, & \text{if } w_j(m_i) = 0 \wedge n_j(m_i) = 0 \\
 w_j(m_i), & \text{if } 0 \leq n_j(m_i) \leq 1 \\
 \dfrac{w_j(m_i)}{n_j(m_i)}, & \text{otherwise.}
\end{cases}
\]

What $\sigma_j(m_i)$ is capturing is the ratio between the observed meme rating $w_j(m_i)$ and the expected meme rating $n_j(m_i)$ given the website popularity in week $j$ and the meme $i$ overall popularity. When we do not observe nor expect any rating, the ratio is undefined ($\frac{0}{0}$), but we are actually observing as much as we expect, therefore we set $\sigma_j(m_i) = 1$. When we expect less than one rating, i.e. $n_j(m_i) \leq 1$, we just deal with it as if we expect just one rating, or $n_j(m_i) = 1$, otherwise $\sigma_j(m_i)$ would tend to $\infty$ even for small $w_j(m_i)$.

In Figure \ref{fig:competition}(b) we represent the normalized meme rating vectors of the two meme rating vectors depicted in Figure \ref{fig:competition}(a). As we can see, the two vectors can now be confronted.

\subsubsection{Rating Vectors Comparison}
We are interested in understanding if the rising of a meme makes the fall of the other meme more likely. For this reason, we calculate a set of conditional probabilities.

Each meme $m$ has a given probability of having $\sigma(m) > 1$, i.e. it is more popular than expected; and $\sigma(m) < 1$, i.e. it is less popular than expected. We refer to these probabilities as $p_m(\sigma(m) > 1)$ and $p_m(\sigma(m) < 1)$. We systematically calculate eight conditional probabilities for each couple of memes $m_i$ and $m_j$: 

\begin{itemize}
\item $p_{m_i}(\sigma(m_i) > 1 | \sigma(m_j) > 1)$;
\item $p_{m_i}(\sigma(m_i) > 1 | \sigma(m_j) < 1)$;
\item $p_{m_i}(\sigma(m_i) < 1 | \sigma(m_j) > 1)$;
\item $p_{m_i}(\sigma(m_i) < 1 | \sigma(m_j) < 1)$.
\end{itemize}

and the reverse conditional probabilities given $\sigma(m_i)$. $p_{m_i}(\sigma(m_i) > 1 | \sigma(m_j) > 1)$ is recording the probability of meme $m_i$ to be more popular than expected given that we observed that meme $m_j$ is more popular than expected; $p_{m_i}(\sigma(m_i) > 1 | \sigma(m_j) < 1)$ is the probability of meme $m_i$ being more popular than expected given that meme $m_j$ is less popular than expected; and so on.

Then, we say that $m_i$ and $m_j$ are in competition if all the following inequalities hold true:

\begin{itemize}
\item $p_{m_i}(\sigma(m_i) > 1) < p_{m_i}(\sigma(m_i) > 1 | \sigma(m_j) < 1)$;
\item $p_{m_i}(\sigma(m_i) < 1) < p_{m_i}(\sigma(m_i) < 1 | \sigma(m_j) > 1)$;
\item $p_{m_j}(\sigma(m_j) > 1) < p_{m_j}(\sigma(m_j) > 1 | \sigma(m_i) < 1)$;
\item $p_{m_j}(\sigma(m_j) < 1) < p_{m_j}(\sigma(m_j) < 1 | \sigma(m_i) > 1)$.
\end{itemize}

On the other hand, $m_i$ and $m_j$ are in collaboration if all the following inequalities holds true:

\begin{itemize}
\item $p_{m_i}(\sigma(m_i) > 1) < p_{m_i}(\sigma(m_i) > 1 | \sigma(m_j) > 1)$;
\item $p_{m_i}(\sigma(m_i) < 1) < p_{m_i}(\sigma(m_i) < 1 | \sigma(m_j) < 1)$;
\item $p_{m_j}(\sigma(m_j) > 1) < p_{m_j}(\sigma(m_j) > 1 | \sigma(m_i) > 1)$;
\item $p_{m_j}(\sigma(m_j) < 1) < p_{m_j}(\sigma(m_j) < 1 | \sigma(m_i) < 1)$.
\end{itemize}

So, if the conditional probability of $m_i$ being more (or less) successful given that $m_j$ was less (or more) successful is higher than the independent probability of $m_i$ being more (or less) successful, and viceversa, then $m_i$ and $m_j$ are said to be in ``competition''. In the opposite case, when the conditional probability of $m_i$ being more (or less) successful given that $m_j$ was more (or less) successful is higher than the independent probability of $m_i$ being more (or less) successful, and viceversa, then $m_i$ and $m_j$ are said to be in ``collaboration''.

We discard competition and collaboration relationships if the two memes were not present together in the Quickmeme website for more than 25 weeks. We verify if the two memes are expected to generate a competition and collaboration relationships just by looking at their popularity and at the Quickmeme website popularity. We use one null matrix as the ``observed'' meme ratings and another null matrix as the ``expected'' values. If two memes rise and fall together in the null matrix, then an observed ``collaboration'' between them may be due solely to external factors. So, if they appear to ``collaborate'', they are in fact only behaving as expected.

For this check we make use of the $N_{m,w}^{1}$ and $N_{m,w}^{2}$ null matrices. We calculate the $\sigma_j(m_i)$ values for the $N_{m,w}^{1}$ matrix with the same procedure described above and using as a null model the $N_{m,w}^{2}$ matrix. Then, we calculate the probability values. If we find competition and collaboration relationships between any two memes in this procedure, we remove that couple of memes from the possible competing or collaborating memes.

Overall, the result of this procedure is a matrix $C_{m,m}$. In $C_{m,m}$ both rows and columns are memes. Each entry of $C_{m,m}$ can take three possible values: 1 if there is collaboration between the two memes, -1 if there is competition and 0 if the two memes are independent from each other.

\subsection{Meme Collaboration}\label{sec:collaboration}
The $C_{m,m}$ matrix is a square matrix $499 \times 499$. Given that the relations are reciprocal, the $C_{m,m}$ matrix is symmetric, and thus there are $\frac{499 \times 498}{2} = 124,251$ possible relationships between memes. Ten of the found collaborations and competitions were also found in the null matrices, thus they were eliminated from $C_{m,m}$. In the end, $18,744$ relationships ($15.08\%$) are equal to $-1$, while $38,619$ relationships ($31.08\%$) are positive. Thus, on average, each meme has a popularity anti-correlation (competition) with $\frac{18,744 \times 2}{499} = 75.13$ memes and it has a popularity correlation (collaboration) with $\frac{38,619 \times 2}{499} = 154.79$ memes. From this fact we can deduct that, among the memes in Quickmeme.com, collaboration is more common than competition. This is expected, as all the meme coexist in the same website and it is very easy to give a positive rating to every meme. For this reason, we first start considering the collaboration phenomena.

\begin{table}
\centering
\small
\begin{tabular}{|l|rrr|}
\hline
Meme & $k_1$ & $k_{-1}$ & $|m|$\\
\hline
College Freshman & 308 & 41 & 974 \\
Jackass Boyfriend & 306 & 34 & 81 \\
Tech Impaired Duck & 304 & 32 & 555\\
All The Things & 304 & 33 & 866 \\
I Got This & 296 & 42 & 107 \\
\hline
Hipster Dog & 36 & 370 & 208 \\
Scumbag Reddit & 64 & 298 & 820\\
Lazy Bachelor Bear & 65 & 280 & 39\\
Guido Jesus & 90 & 270 & 211\\
Scumbag Parents & 88 & 268 & 675\\
\hline
\end{tabular}
\caption{The top five memes in collaboration (above half) and competition (bottom half).}
\label{tab:collaborators-competitors}
\end{table}

We report in Table \ref{tab:collaborators-competitors} the top 5 memes with positive (top half of the table) and negative (bottom half of the table) popularity correlations. Column $k_1$ reports the number of memes correlated with meme $m$, column $k_{-1}$ the number of memes anti-correlated with $m$ and $|m|$ is the number of meme implementations. The average popularity in number of meme implementations $|m|$ of the top meme collaborators is higher than the average popularity of the top meme competitors. Our idea is that competition generates cluster of correlated memes, that we can call ``meme organisms''.

\begin{figure}
\centering
\includegraphics[width=.95\columnwidth]{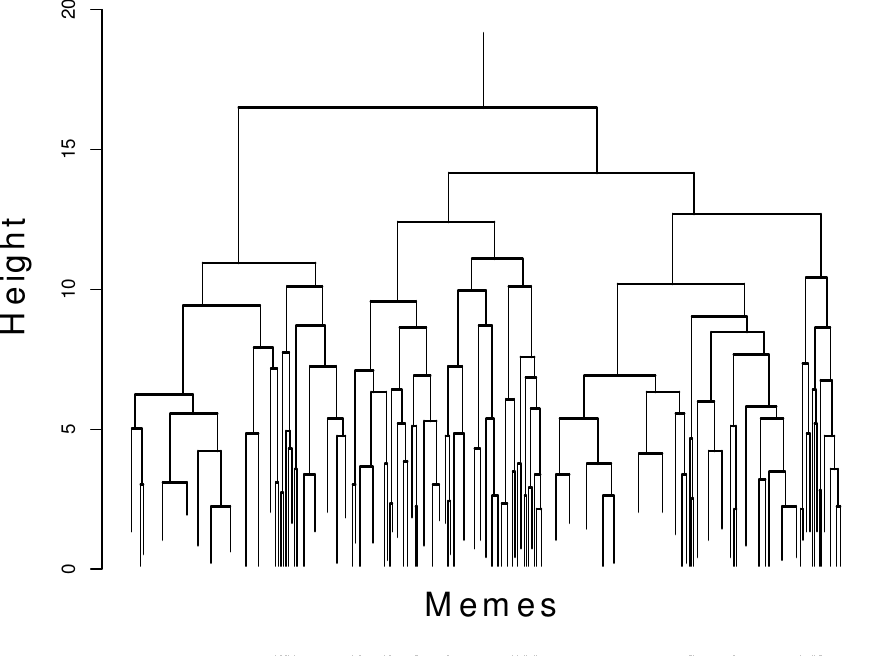}
\caption{The result of the hierarchical cluster of $C_{m,m}$ matrix. For clarity, we show only the top part of the dendrogram.}
\label{fig:meme-organisms}
\end{figure}

To understand if there really are meme organisms, we need to understand if large groups of memes tend to correlate to each other. This intuition corresponds to performing clustering on the $C_{m,m}$ matrix. To this end, we remove from it all the $-1$ entries, which we consider as $0$. Then, we perform a hierarchical clustering\footnote{We used the hclust function in the R statistical software.}, using the complete linkage method to calculate how distant a couple of memes, or a couple of meme clusters, are from each other. In complete linkage, the distance between two clusters is computed as the maximum distance between a pair of objects, one in one cluster, and one in the other. By being very demanding on the similarity of memes, we make sure that we group in each cluster only very similar memes.

We report in Figure \ref{fig:meme-organisms} the dendrogram of the resulting clustering of the $C_{m,m}$ matrix. As we can see, overall the cluster structure of the $C_{m,m}$ matrix shows us that the matrix itself has a modular structure, as there are well defined clusters. To obtain the clusters we need to define a distance threshold below which all memes are grouped. Different thresholds can be chosen and we leave as a future work the task of defining the best one to obtain the meme organisms. Here, we chose an arbitrary reasonable cut height and we call ``organisms'' the resulting meme clusters.

\begin{figure}
\centering
\subfloat[Chemistry Cat]{\includegraphics[width=.35\columnwidth]{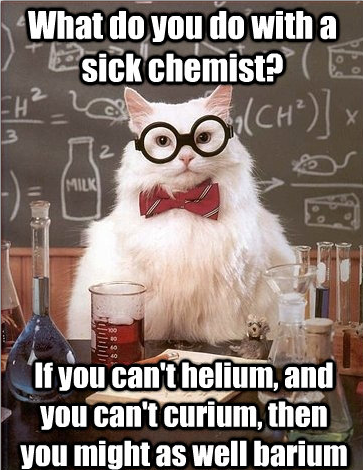}}\quad
\subfloat[Dwight]{\includegraphics[width=.6\columnwidth]{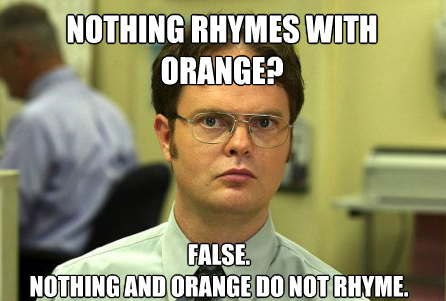}}
\caption{The meme cluster \#45.}
\label{fig:meme-collaborators}
\end{figure}

As an illustration, we present an example of one memes organism. For clarity, out of the 103 clusters (composed by $4.8$ memes on average) we chose to show cluster \#45, composed by two memes. The two memes are ``Chemistry Cat'' and ``Dwight'', and we show a meme implementation of both in Figure \ref{fig:meme-collaborators}. Chemistry Cat is a meme used to make puns using scientific concepts, while Dwight is a character of a popular TV show, who is used to make pedantic remarks about common knowledge. The two organisms are intuitively part of very related geeky humor.

\begin{table}
\centering
\small
\begin{tabular}{|l|rr|}
\hline
Probability & Chemistry Cat & Dwight\\
\hline
$p_{m_i}(\sigma(m_i) > 1)$ & $20.75\%$ & $18.87\%$ \\
$p_{m_i}(\sigma(m_i) < 1)$ & $79.25\%$ & $81.13\%$ \\
$p_{m_i}(\sigma(m_i) > 1 | \sigma(m_j) > 1)$ & $60.00\%$ & $54.55\%$ \\
$p_{m_i}(\sigma(m_i) > 1 | \sigma(m_j) < 1)$ & $11.63\%$ & $9.52\%$ \\
$p_{m_i}(\sigma(m_i) < 1 | \sigma(m_j) > 1)$ & $40.00\%$ & $45.45\%$ \\
$p_{m_i}(\sigma(m_i) < 1 | \sigma(m_j) < 1)$ & $88.37\%$ & $90.48\%$ \\
\hline
\end{tabular}
\caption{The probabilities of the memes in cluster \#45 of being successful and unsuccessful.}
\label{tab:meme-collaborators}
\end{table}

In Table \ref{tab:meme-collaborators} we provide the process through which we say that the two memes are collaborating: their independent and conditional probabilities of being successful and unsuccessful. Chemistry Cat is successful on $20.75\%$ of the weeks, but if also Dwight is successful its odds raise to $60\%$, while if Dwight is unsuccessful its odds lower to $11.63\%$. The odds of being unsuccessful raise when Dwight is unsuccessful. The inequalities hold also for the Dwight meme.

\subsection{Meme Competition}\label{sec:examples}

\begin{figure}
\centering
\subfloat[Against ``First World Problems'']{\includegraphics[width=.475\columnwidth]{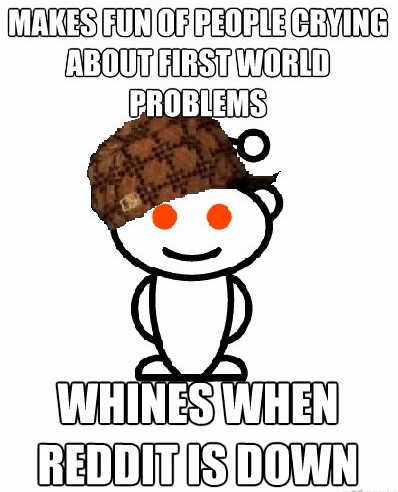}}\quad
\subfloat[Against ``Hipster'']{\includegraphics[width=.475\columnwidth]{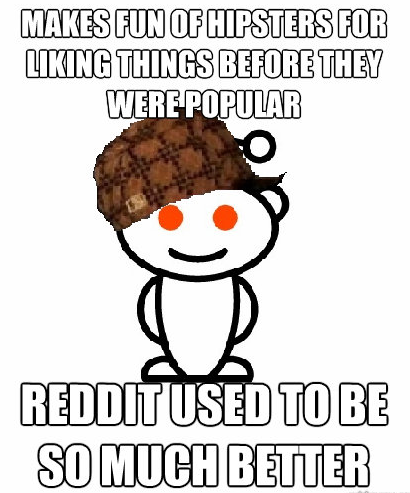}}
\caption{Two meme implementations of Scumbag Reddit meme.}
\label{fig:meme-competitors}
\end{figure}

While less widespread than collaboration, there is also competition in the meme pool: some memes have a disproportionate amount of anti-correlations with other memes.

In this case, one of the most evident explanation can be found in the way some users use the meme itself. An example of this is reported in Figure \ref{fig:meme-competitors}. Figure \ref{fig:meme-competitors} reports two meme implementations of the meme ``Scumbag Reddit'' (SR), a meme used for self-critique of many widespread user behaviors on the popular social bookmarking website Reddit.com (whose users make heavy use of Quickmeme.com). As we saw in Table \ref{tab:collaborators-competitors}, SR meme is the second most competitive meme. From Figure \ref{fig:meme-competitors} we can understand why: often users adopt this meme to state something about other memes (both implementations refer to other memes, specifically Figure \ref{fig:meme-competitors}(a) refers to ``First World Problems'' memes and Figure \ref{fig:meme-competitors}(b) refers to ``Hipster'' memes).

\begin{table}
\scriptsize
\centering
\begin{tabular}{|l|rr|}
\hline
Probability & Scumbag Reddit & FWP Cat\\
\hline
$p_{m_i}(\sigma(m_i) > 1)$ & $26.42\%$ & $11.32\%$ \\
$p_{m_i}(\sigma(m_i) < 1)$ & $73.58\%$ & $88.68\%$ \\
$p_{m_i}(\sigma(m_i) > 1 | \sigma(m_j) > 1)$ & $16.66\%$ & $7.14\%$ \\
$p_{m_i}(\sigma(m_i) > 1 | \sigma(m_j) < 1)$ & $27.66\%$ & $12.82\%$ \\
$p_{m_i}(\sigma(m_i) < 1 | \sigma(m_j) > 1)$ & $83.33\%$ & $92.86\%$ \\
$p_{m_i}(\sigma(m_i) < 1 | \sigma(m_j) < 1)$ & $72.34\%$ & $87.18\%$ \\
\hline
\multicolumn{1}{r}{} & \multicolumn{1}{r}{} & \multicolumn{1}{r}{} \\
\hline
Probability & Scumbag Reddit & Hipster Barista\\
\hline
$p_{m_i}(\sigma(m_i) > 1)$ & $26.42\%$ & $39.62\%$ \\
$p_{m_i}(\sigma(m_i) < 1)$ & $73.58\%$ & $60.38\%$ \\
$p_{m_i}(\sigma(m_i) > 1 | \sigma(m_j) > 1)$ & $9.52\%$ & $14.29\%$ \\
$p_{m_i}(\sigma(m_i) > 1 | \sigma(m_j) < 1)$ & $37.50\%$ & $48.72\%$ \\
$p_{m_i}(\sigma(m_i) < 1 | \sigma(m_j) > 1)$ & $90.48\%$ & $85.71\%$ \\
$p_{m_i}(\sigma(m_i) < 1 | \sigma(m_j) < 1)$ & $62.50\%$ & $51.28\%$ \\
\hline
\end{tabular}
\caption{The probabilities of Scumbag Reddit meme being successful and unsuccessful given an example of First World Problems and Hipster memes.}
\label{tab:meme-competitors}
\end{table}

To prove this point, we report in Table \ref{tab:meme-competitors} the conditional probabilities of SR meme with a First World Problem (FWP) meme and one Hipster meme. The fact that SR meme is popular negatively influences the success odds of the other two memes and viceversa.

\section{Meme Success}\label{sec:success}
In this section we study the characteristics of successful and unsuccessful memes, aiming at a description of the characteristics that are correlated with meme success. First, we need to define the set of meme features we want to study. Then, we establish a criterion to determine if a meme is successful or unsuccessful. Finally, we use a decision tree algorithm to describe how the features make the probability of one meme to be successful higher or lower.

\subsection{Meme Features}
We decided to focus on four main features of the memes. The features are: number of collaborators, number of competitors, the fact that a meme is in a meme organism and the entity of its popularity peak over its average popularity.

The number of collaborators and the number of competitors are how many memes are recorded in competition and in collaboration according to the procedure described in Section \ref{sec:nullmodel}. Instead of taking the absolute number, which may generate a decision tree too deep and difficult to interpret, we decided to create three bins for each of these two features. Memes are classified as highly competitive/collaborative, average competitive/collaborative or lowly competitive/collaborative. We defined the thresholds to classify a meme into one of these three categories in order to have balanced, i.e. equally populated, bins for high and low competitors/collaborators.

\begin{figure*}
\centering
\subfloat[Competitors]{\includegraphics[width=.95\columnwidth]{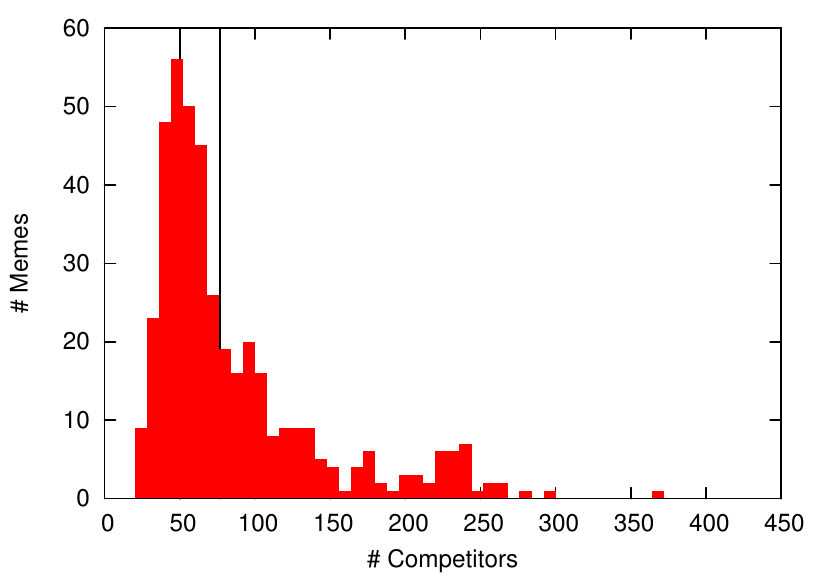}}\quad
\subfloat[Collaborators]{\includegraphics[width=.95\columnwidth]{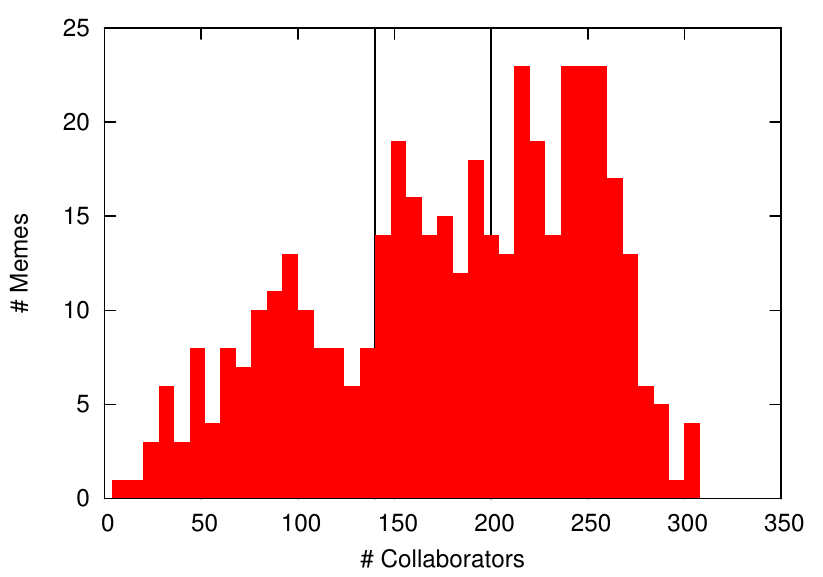}}
\caption{Distributions per meme.}
\label{fig:compcoll-distr}
\end{figure*}

Figures \ref{fig:compcoll-distr} (a-b) depict the distributions for each meme of the number of competitors and collaborators, respectively. The black lines indicate our thresholds for the bins. We also sum up the values of the thresholds and how many memes fell into that particular bin in Table \ref{tab:dtree-char}. To be highly competitive, a meme needs to have more than $77$ competitors (and $172$ memes satisfy this constraint), while to be low competitive it is required to have less than $50$ competitors, with $174$ memes falling into this bin.

The third feature records whether a meme is part of a meme organism or not. We introduced the general methodology to detect meme organisms in Section \ref{sec:collaboration}. Here, we make use of the organisms extracted in that section. To be in an organism, a meme is required to be present in a cluster with at least other two memes, given our definition of organisms in Section \ref{sec:collaboration}. Table \ref{tab:dtree-char} reports the number of memes being in one organism and the ones that are not.

%

\begin{table}
\centering
\scriptsize
\begin{tabular}{|ll|rrr|}
\hline
\multicolumn{2}{|l|}{Feature} & $> AVG$ & $= AVG$ & $< AVG$\\
\hline
\multirow{2}{*}{Competition} & Threshold & $x > 77$ & $50 \leq x \leq 77$ & $x < 50$\\
& \# Memes & $172$ & $153$ & $174$\\
\hline
\multirow{2}{*}{Collaboration} & Threshold & $x > 200$ & $140 \leq x \leq 200$ & $x < 140$\\
& \# Memes & $195$ & $116$ & $188$\\
\hline
\multirow{2}{*}{In Organism} & Threshold & $True$ & N/A & $False$\\
& \# Memes & $336$ & N/A & $163$\\
\hline
\multirow{2}{*}{Peak} & Threshold & $x > 25$ & N/A & $x \leq 25$\\
& \# Memes & $231$ & N/A & $268$\\
\hline
\multirow{2}{*}{Successful} & Threshold & $True$ & N/A & $False$\\
& \# Memes & $177$ & N/A & $322$\\
\hline
\end{tabular}
\caption{The thresholds and number of memes for our feature bins.}
\label{tab:dtree-char}
\end{table}

\begin{figure}
\centering
\includegraphics[width=.95\columnwidth]{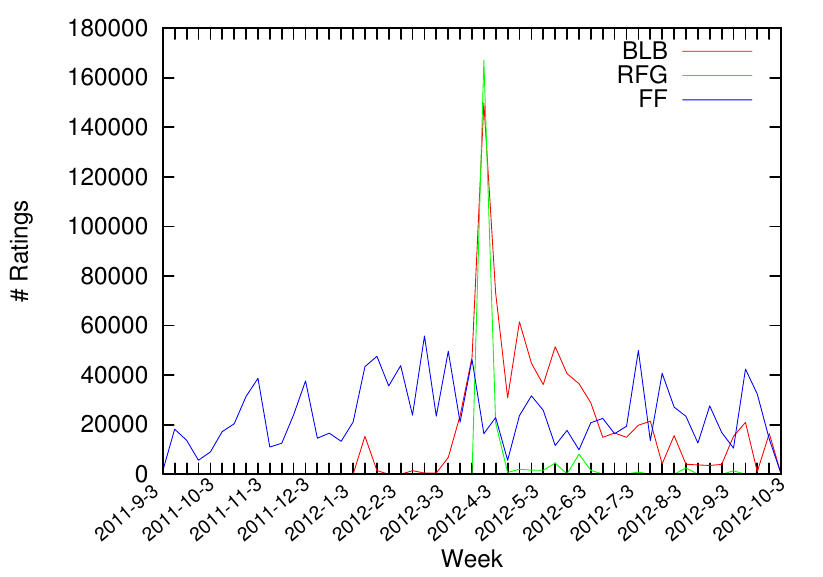}
\caption{Some meme rating vectors with different peaks.}
\label{fig:pop-peak}
\end{figure}

The last feature is the relative height of the popularity peak of a meme over its average popularity. Only a handful of memes are constantly popular on Quickmeme. Often, memes have popularity peaks. These peaks do not necessarily happen when a meme is born, but when a user creates a very successful meme implementation of it and this implementation hits the front page. Then, many users produce all possible variants of this meme implementation in a limited time span and many of these usually hit the front page too\footnote{It is the so-called ``Karma train'' and there is a meme that signals it: \url{http://www.quickmeme.com/Mad-Karma-with-Jim-Cramer/?upcoming}}. At this point, either the meme stays popular or it fades back into oblivion. Figure \ref{fig:pop-peak} depicts the meme rating vectors of three different memes. ``Bad Luck Brian'' (BLB) has a peak several weeks after its creation and it manages to stay somewhat popular after its large peak. ``Ridiculously Photogenic Guy'' (RFG), instead, has an immediate peak larger than BLB, but it then fades into oblivion as an ephemeral trend. Finally, ``Futurama Fry'' (FF) has no large popularity peak at all, but it lives as successful meme.

To calculate the last feature we take the amount of ratings a meme got in its most popular week and we divide it with the average weekly rating of that meme. Like we did for collaborator and competitor features, we create equipopulated bins for this feature. In this case, we create two bins: ``Above average'' for memes which have a popularity peak 25 times or more higher than their average popularity, and ``Below Average'' if the popularity peak is lower than 25 times the average popularity. The peak values for the memes reported in Figure \ref{fig:pop-peak} are 9.6 for BLB, 41.44 for RPG and 2.36 for FF. We report the number of memes in both bins in Table \ref{tab:dtree-char}.

\subsection{Measuring Success}
\begin{figure}
\centering
\includegraphics[width=.95\columnwidth]{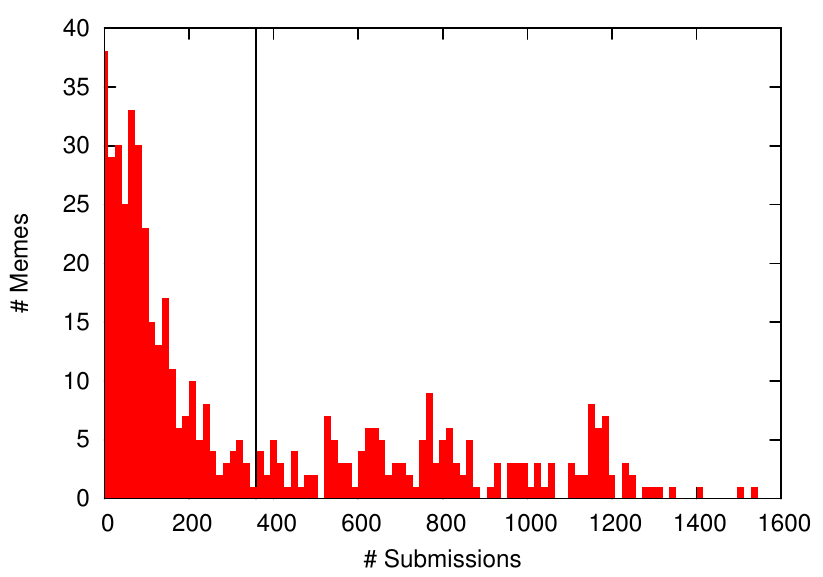}
\caption{Distribution of the number of submissions per meme.}
\label{fig:sub-distr}
\end{figure}

There are many alternatives to measure how successful a meme is. Since we already used a proxy of the number of ratings a meme gets as one of the meme features (the popularity peak), we cannot use the total number of ratings of a meme as a measure. For this reason, we look instead at the number of implementations a meme gets: the more implementations the more we can say that a meme is persistent in people's mind. For each meme $m_i$ (the set of all of its implementations) we check $|m_i|$: if it is higher than the average $|m|$ then we consider the meme successful, otherwise it is unsuccessful. Since we have $178,801$ meme implementations distributed in $499$ memes (see Section \ref{sec:data}), our threshold equals to $358$ implementations. Figure \ref{fig:sub-distr} depicts the distribution of the number of implementations per meme (the black line represents our threshold), and Table \ref{tab:dtree-char} reports how many memes are (un)successful.

As a result, each meme is now described as a list of features. For example, the ``Socially Awkward Penguin'' meme presented in Section \ref{sec:data} has $1,413$ meme implementations making it a successful meme. SAP also has: $48$ competitors, $285$ collaborators, a popularity peak $3.91\times$ its average popularity and it is in an organism with $11$ memes. Therefore, according to Table \ref{tab:dtree-char}: $\{Competition:\ <AVG$, $Collaboration:\ >AVG$, $In\ Organism:\ True$, $Peak:\ <AVG$, $Successful:\ True\}$. 

\subsection{Extracting and Interpreting the Decision Tree}
Each meme generates a record with the procedure described in the previous sections. The records are then used as input to a decision tree algorithm. As the decision tree's target variable we used ``Success''. We used the decision tree implementation described in \cite{dtree}. In \cite{dtree}, the tree can be pruned if it contains too many levels. We also decided to merge some leaves of the tree to facilitate its interpretation.

\begin{figure*}
\centering
\includegraphics[width=1.9\columnwidth]{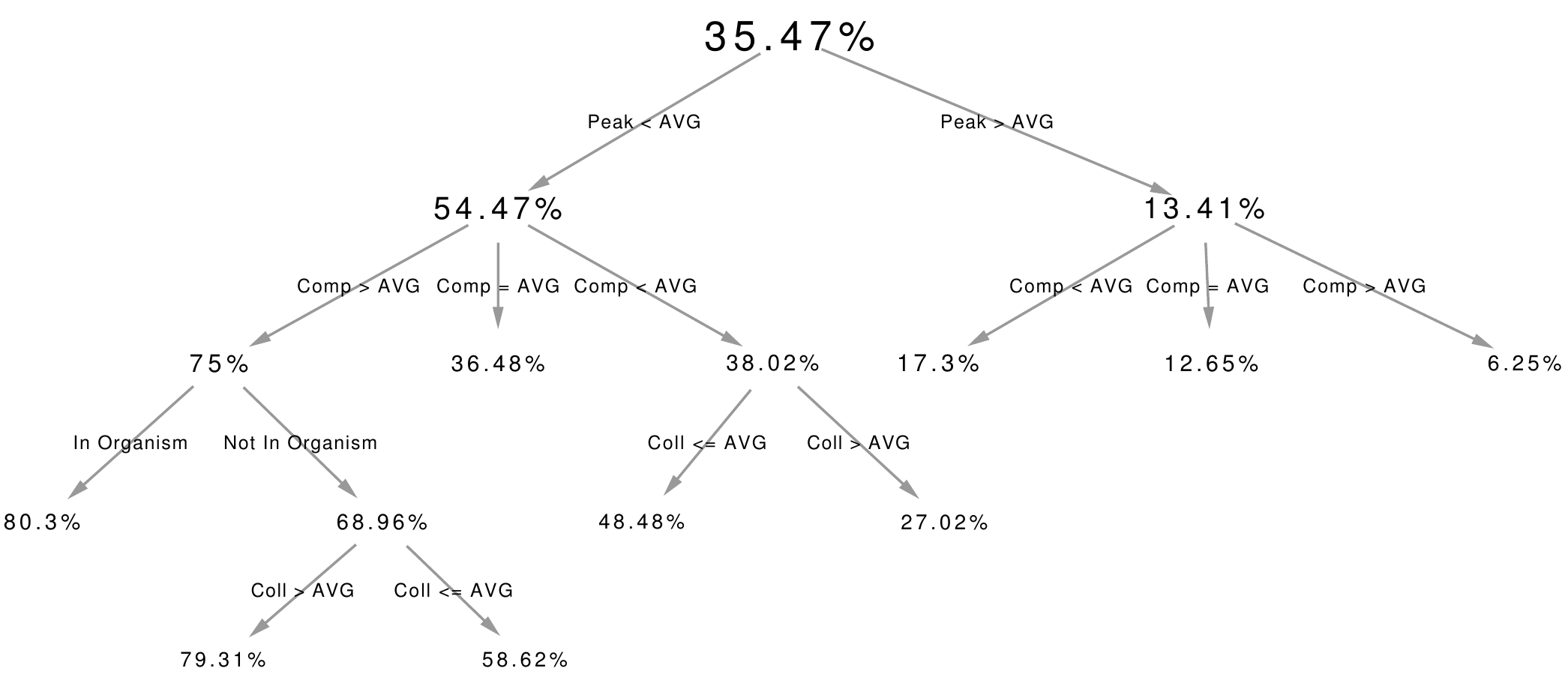}
\caption{The decision tree describing the success odds given the meme characteristics.}
\label{fig:success-tree}
\end{figure*}

Our final decision tree is depicted in Figure \ref{fig:success-tree}. In Figure \ref{fig:success-tree}, the nodes of our tree contain the share of memes in the particular node tree that are successful. In the root of the tree (the top node in the upper part of the figure), we report the baseline probability of a meme being successful. As we see in Table \ref{tab:dtree-char}, there are $177$ successful memes in our dataset. Therefore, the root node reports $\frac{177}{499} \times 100 = 35.47\%$.

Then, each node is split in two or more branches, according to the feature that best separates successful from unsuccessful memes. The most important feature is ``Peak'': if there was a high popularity peak then the odds of being successful lowers to $13.41\%$. Memes with low popularity peaks have a success probability of $54.47\%$.

In the two resulting branches we can observe an interesting fact. High popularity peaks make the number of competitors negatively correlated to the odds of being successful: a large or average number of competitors make the successful odds drop to $6.25\%$ and to $12.65\%$, while a small amount of competitors make the meme slightly more likely to be successful ($17.3\%$). On the other branch, i.e. for memes without high popularity peaks, the number of competitors is actually positively correlated with success odds. Highly competitive memes have $75\%$ chances of being successful, more than average competitors ($36.48\%$) and low competitors ($38.02\%$). A possible explanation of this fact may be the following: if there is a popularity peak, then a meme will be used frequently only if it is not too similar to many other memes. If, instead, there is no popularity peak, users are likely to stick with it, and keep being used together with other memes, thus competing with them.

If a meme has no peak and low number of competitors, a higher than average number of collaborators correlates to its success odds (we merged the two bins because the distinction between them was not significant).

On the other hand, if a meme has no peak and a high number of competitors, the best correlation is registered in it being in a coherent meme organism. Memes in this situation have increased odds of being successful (equals to $80.3\%$). If the meme is not in an organism, having a high number of collaborators still highly correlates with the success rate ($79.31\%$), while an average or lower than average number of collaborators decrease success odds to $58.62\%$.

To sum up: competition is anti-correlated with the odds of being successful if a meme also happen to have experienced popularity peaks. If it did not, it is a good thing only if the meme is part of a meme organism or at least it can count on many collaborations with other memes. Being a collaborative meme correlates with success.

\section{Conclusion and Future Works}\label{sec:conclusions}
In this paper we presented an empirical approach to the study of memes, by analyzing a controlled dataset focusing on a particular class of internet memes. As opposed to the main approach in literature, which studies the characteristics of social networks in favoring meme spreads, we proposed a perspective without the use of network effects, as we think that studying the characteristics of memes can provide useful insights on cultural patterns, as opposed to the social patterns studied with networks.

We studied the behavior of internet memes in the website Quickmeme.com. We proved that in Quickmeme there are actual memes as they compete and collaborate, sometimes clustering in larger ensembles. We showed as different meme characteristics are associated with increased or decreased odds for the meme of being popular.

Our work paves the way to a number of future developments. First, the network approach is not necessarily mutually exclusive with our work. Combining the meme competing studies in networks with our approach may provide useful insights about meme dynamics. Also, our approach in analyzing the temporal evolution of memes is still not perfect: for example it considers each snapshot in a meme's timeline as equally important, whereas a more dynamic approach, such the one presented in \cite{pakdd10}, can split it in different eras with distinct characteristics.

We believe that this work can be part of and increased understanding about how memes work, with the hope of shredding more light on the complex dynamics of human cultural patterns.\\

\textbf{Acknowledgments}. The author wants to thank Clara Vandeweerdt and Muhammed Yildirim for useful discussions and feedback. 

\bibliographystyle{aaai}

\bibliography{biblio}

\balance

\end{document}